\def\arxivversion{1}
\documentclass[letterpaper]{article} 

\newif\ifarxivpreprint
\ifdefined\arxivversion
  \arxivpreprinttrue
\else
  \arxivpreprintfalse
\fi

\newif\ifanonymous
\anonymousfalse
\ifanonymous
  \usepackage[submission]{aaai2027}
\else
  \usepackage{aaai2027}
\fi
\ifarxivpreprint
  \nocopyright
\fi

\usepackage[hyphens]{url} 
\usepackage{graphicx} 
\usepackage{natbib} 
\usepackage{caption} 
\usepackage{booktabs}
\usepackage{amsmath,amssymb}
\usepackage{array}
\usepackage{multirow}
\usepackage{xcolor}
\usepackage{tabularx}

\ifdefined\arxivversion
\else
\fi

\newcommand{\closer}{\textsc{Closer-Bench}}
\newcommand{\specToRTL}{spec$\rightarrow$RTL}
\newcommand{\rtlToGDS}{RTL$\rightarrow$GDS}
\newcommand{\specToGDS}{spec$\rightarrow$GDS}
\newcommand{\best}{\textbf}

\title{CLOSER-Bench: Evaluating Budgeted Cross-Stage Design Closure for Hardware Agents}

\author{
Peilong Zhou\textsuperscript{\rm 1,\rm 2,\rm 3},
Zhirong Chen\textsuperscript{\rm 1,\rm 2},
Cangyuan Li\textsuperscript{\rm 1,\rm 2},
Haoyu Gao\textsuperscript{\rm 1,\rm 2,\rm 3},
\\
Kaiyan Chang\textsuperscript{\rm 1,\rm 2},
Ziming Qu\textsuperscript{\rm 1,\rm 2},
Ying Wang\textsuperscript{\rm 1,\rm 2}\corresponding
}

\affiliations{
\textsuperscript{\rm 1}SKLP, Institute of Computing Technology, Chinese Academy of Sciences\\
\textsuperscript{\rm 2}University of the Chinese Academy of Sciences\\
\textsuperscript{\rm 3}School of Advanced Interdisciplinary Sciences\\
wangying2009@ict.ac.cn
}

\begin{document}

\maketitle

\begin{abstract}
Hardware engineering exposes coding agents to a form of long-horizon work that
is difficult to capture with pass-at-$k$: progress is continuous, tool feedback
is delayed and heterogeneous, and a backend failure may require revising RTL
rather than tuning another physical-design parameter.  Existing benchmarks
measure RTL generation, repository repair, verification, PPA evolution, or
physical implementation, but their different designs and oracles make it hard
to determine where an agent succeeds or fails across abstraction boundaries.
We introduce \closer{}, a controlled evaluation protocol for budgeted
cross-stage design closure.  For one design and one hidden objective, it pairs
\specToRTL{}, \rtlToGDS{}, and \specToGDS{} tasks, records every simulator,
synthesis, STA, and place-and-route invocation, and measures final quality,
anytime progress, tool cost, and cross-stage recovery.  The benchmark is built
on open-source Verilator, Yosys, OpenROAD, KLayout, Sky130, and the Harbor agent
harness.  A ten-task pilot spanning RTL repair, mutation-based verification,
coverage, PPA optimization, design-space exploration, cross-model debugging,
and security establishes the executable harness and exposes a sharp
completion--closure gap: three agents solve a localized AXI repair task, while
the matched verification-closure task separates a frontier agent from two
otherwise successful baselines.  We further validate a full RTL-to-GDS flow
and construct a macro-based AXI/DMA streaming accelerator for the stage-paired
evaluation.  These results motivate treating hardware closure as a budgeted
sequential decision problem rather than a collection of independent code
generation tasks.
\end{abstract}

\section{Introduction}

Recent hardware benchmarks show that large language models can generate RTL,
repair repository-level bugs, write verification artifacts, and interact with
electronic-design-automation (EDA) tools
\cite{liu2023verilogeval,lu2024rtllm,pinckney2025cvdp,cui2026hwe,zou2026phoenix}.
The frontier is consequently moving from one-shot functional correctness to
\emph{closure}: repeatedly improving a design until functional, timing,
physical, and resource constraints hold together.  Closure is naturally
long-horizon.  A simulation takes seconds, synthesis and static timing analysis
(STA) take minutes, and a place-and-route (PnR) attempt can take tens of minutes.
Worse, feedback is only partially localizing.  Negative slack after routing may
be caused by RTL microarchitecture, an incorrect constraint, macro placement,
or flow parameters.  A capable agent must decide both \emph{what} to change and
\emph{which abstraction layer} to revisit.

Current evaluations do not isolate this ability.  VerilogEval and RTLLM begin
from specifications and largely stop at simulation or synthesis
\cite{liu2023verilogeval,lu2024rtllm}.  CVDP, ChipBench, HWE-Bench, and
Phoenix-bench broaden front-end and repository-level evaluation
\cite{pinckney2025cvdp,yu2026chipbench,cui2026hwe,zou2026phoenix}, while CktEvo
studies function-preserving RTL PPA evolution \cite{shi2026cktevo}.  At the
backend, ChatEDA, ORFS-agent, ASIC-Agent, and PDAGENT-BENCH execute or optimize
physical-design workflows
\cite{he2024chateda,ghose2025orfs,allam2025asic,li2026pdagent}.  Design
Conductor goes further and demonstrates autonomous specification-to-GDSII CPU
design \cite{verkor2026conductor}.  These are substantial advances, but they do
not provide a controlled answer to a basic question: if the same agent is
successful at the front end and backend separately, will it allocate feedback
and recovery actions correctly when the boundary is removed?

We propose \closer{} (\textbf{C}ross-\textbf{L}ayer
\textbf{O}bjective-driven \textbf{S}tage-paired \textbf{E}valuation under
\textbf{R}esource constraints).  Its central experimental control is to derive
three tasks from the same design specification and hidden objective:
\specToRTL{} measures front-end completion, \rtlToGDS{} measures backend
closure from a functionally correct but physically weak implementation, and
\specToGDS{} exposes the full cross-stage action space.  Shared workloads,
constraints, and signoff checks make performance differences attributable to
stage boundaries rather than unrelated task content.

\paragraph{Research questions.}
\textbf{RQ1:} Do explicit stage boundaries improve reliability and efficiency?
\textbf{RQ2:} When downstream feedback invalidates the current approach, can an
agent recover at the correct abstraction layer?  \textbf{RQ3:} How do final
quality and failure modes change as expensive-tool and wall-clock budgets scale?

\paragraph{Contributions.}
This paper makes three contributions.  First, it defines a stage-paired,
budgeted protocol for cross-stage hardware closure, including unified trajectory
and anytime metrics.  Second, it implements the protocol with open-source EDA
tools and a shared AXI/DMA streaming design with an SRAM macro and executable
functional oracle.  Third, a ten-task pilot and physical-flow audit identify
which existing tasks measure sustained closure and which merely add expensive
runtime to a locally solvable problem.  We report the completed pilot evidence
and treat the full three-way agent matrix as future evaluation rather than
estimating unobserved results.

\begin{figure*}[t]
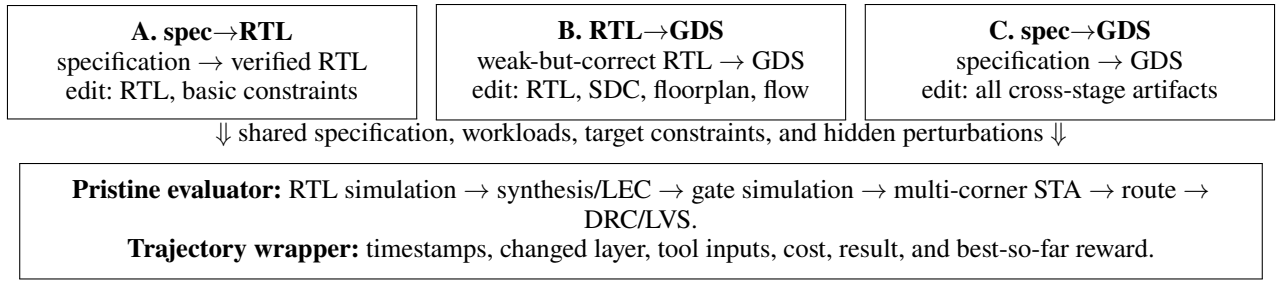

\centering
\setlength{\fboxsep}{6pt}
\begin{tabular}{c@{\hspace{0.7em}}c@{\hspace{0.7em}}c}
\fbox{\begin{minipage}{0.28\textwidth}\centering
\textbf{A. \specToRTL{}}\\
specification $\rightarrow$ verified RTL\\
edit: RTL, basic constraints
\end{minipage}}
&
\fbox{\begin{minipage}{0.28\textwidth}\centering
\textbf{B. \rtlToGDS{}}\\
weak-but-correct RTL $\rightarrow$ GDS\\
edit: RTL, SDC, floorplan, flow
\end{minipage}}
&
\fbox{\begin{minipage}{0.28\textwidth}\centering
\textbf{C. \specToGDS{}}\\
specification $\rightarrow$ GDS\\
edit: all cross-stage artifacts
\end{minipage}}
\\[0.8em]
\multicolumn{3}{c}{$\Downarrow$ shared specification, workloads, target constraints, and hidden perturbations $\Downarrow$}\\[0.5em]
\multicolumn{3}{c}{\fbox{\begin{minipage}{0.90\textwidth}\centering
\textbf{Pristine evaluator:} RTL simulation $\rightarrow$ synthesis/LEC $\rightarrow$
gate simulation $\rightarrow$ multi-corner STA $\rightarrow$ route $\rightarrow$ DRC/LVS.\\
\textbf{Trajectory wrapper:} timestamps, changed layer, tool inputs, cost, result, and best-so-far reward.
\end{minipage}}}
\end{tabular}
\caption{The controlled stage-paired design of \closer{}. Task A's oracle is the
front-end projection of the final signoff oracle; Tasks B and C share the full
oracle.  Comparing $A{+}B$ with C separates stage competence from cross-stage
coordination.}
\label{fig:overview}
\end{figure*}

\section{Related Work}

\paragraph{RTL generation, verification, and repair.}
VerilogEval and RTLLM established executable evaluation for natural-language
RTL generation \cite{liu2023verilogeval,lu2024rtllm}; OpenLLM-RTL and CVDP
expanded task scale and diversity \cite{liu2025openllmrtl,pinckney2025cvdp}.
ChipBench includes hierarchical generation, debugging, and reference-model
construction \cite{yu2026chipbench}.  Repository repair is now directly tested
by HWE-Bench's 417 pull-request-derived tasks and Phoenix-bench's 511 Verilator
instances \cite{cui2026hwe,zou2026phoenix}.  HORIZON frames hardware work as
repository-level code evolution and evaluates a hands-free loop across several
of these suites \cite{yu2026horizon}.  Our goal is not another claim of first
repository-scale hardware repair.  Instead, repair and verification are useful
matched controls: closing an open-ended objective can be much harder than
recovering a known completion target on the same IP.

\paragraph{PPA and physical-design agents.}
ChatEDA executes RTL-to-GDSII task decompositions \cite{he2024chateda};
ASIC-Agent connects RTL generation, verification, OpenLane hardening, and chip
integration \cite{allam2025asic}.  CktEvo evaluates repository-level,
function-preserving PPA improvements at RTL \cite{shi2026cktevo}.  ORFS-agent
tunes OpenROAD-flow-scripts parameters and reports resource-efficient
optimization trajectories \cite{ghose2025orfs}.  PDAGENT-BENCH provides 353
tasks and full-flow implementation cases across the physical-design stack
\cite{li2026pdagent}, while Design Conductor demonstrates that a frontier agent
can autonomously produce a timing-clean CPU layout from a short specification
\cite{verkor2026conductor}.  HighTide supplies open, maintainable RTL-to-GDS
benchmark infrastructure \cite{goldblatt2026hightide}.  \closer{} is
complementary: it keeps the RTL edit door open, controls the starting boundary,
and scores the allocation of a limited tool budget rather than only task
completion or final PPA.

\paragraph{End-to-end evaluation.}
HSCO-Bench evaluates agent-driven hardware--software co-design through FPGA
deployment \cite{tsai2026hsco}.  This reinforces a broader lesson: concatenating
stages creates qualitatively different failure modes.  Our stage pairing turns
that observation into a controlled comparison, using identical design intent
and signoff conditions across task boundaries.

\section{Benchmark Design}

\subsection{Design Principles}

\textbf{First feasible is not finished.} Each task exposes a graded objective
after validity gates.  An agent can obtain an early feasible design, then trade
area, delay, power, or verification strength under a remaining budget.
\textbf{Delay must carry information.} Expensive tools reveal different facts
from fast proxies; merely sleeping or rebuilding a large repository does not
make a task long-horizon.  \textbf{Public feedback is not the final oracle.}
Hidden workloads, configurations, corners, and floorplan seeds test whether a
strategy generalizes.  \textbf{Every result has provenance.} Agent failure,
agent timeout with a valid patch, verifier failure, and infrastructure failure
are distinct run states.

\subsection{Stage-Paired Tasks}

Let $s$ denote a common design specification and $\Omega$ a set of hidden
conditions (workload, interface configuration, PVT corner, and floorplan seed).
Task A begins from $s$ and ends at a verified, synthesizable RTL checkpoint.
Task B begins from a canonical implementation $x_B$ that is functionally
correct but intentionally misses one or more physical targets.  Task C begins
from $s$ and allows the union of A and B's edit surfaces.  The functional subset
of $\Omega$ is shared by all tasks; B and C are evaluated by the identical
signoff oracle.  A is evaluated by the oracle projected through synthesis.

The comparison is deliberately not ``three unrelated scores.''  We compare
(i) A against C at the verified-RTL checkpoint, (ii) B against C after the first
feasible layout, and (iii) the sequential composition of independently solved
A and B against C under an equal aggregate budget.  This reveals error
propagation and the cost of deciding when to cross or revisit a stage boundary.

\subsection{Validity and Quality}

The evaluator computes a validity gate
\begin{equation}
G(x,\omega)=G_{\mathrm{func}}G_{\mathrm{lec}}G_{\mathrm{gls}}
G_{\mathrm{sta}}G_{\mathrm{route}}G_{\mathrm{drc}}G_{\mathrm{lvs}},
\end{equation}
where each factor is binary and stage-inapplicable factors equal one.  A
physically feasible candidate is scored against pre-registered targets
$m_j^*$ (e.g., area, delay, power, and cycles):
\begin{equation}
\begin{aligned}
\ell_j(x,\omega) &= \max\!\left\{0,\log\frac{m_j(x,\omega)}{m_j^*}\right\},\\
R(x,\omega) &= G(x,\omega)\exp\!\left[-\sum_j w_j\ell_j(x,\omega)\right].
\end{aligned}
\label{eq:reward}
\end{equation}
Thus invalid shortcuts receive zero, improvements toward any unmet target are
continuous, and meeting all targets yields one.  Raw metrics and the Pareto set
are always reported to expose saturation or an ill-chosen target.

\subsection{Budget and Trajectory Protocol}

The wrapper, rather than the agent, counts calls to four tool classes:
simulation, synthesis/STA, full PnR, and final verifier.  It also records
wall-clock time, model tokens when available, modified files, inferred edit
layer (RTL/constraint/floorplan/flow), tool outcome, and best-so-far reward.
Provisional budgets derived from Phase-0 measurements are shown in
Table~\ref{tab:budget}; every call remains recorded even when nominally
unlimited.

\begin{table}[t]
\centering
\small
\begin{tabular}{lrrr}
\toprule
Resource & Low & Medium & High \\
\midrule
Simulation calls & 40 & 100 & uncapped \\
Synthesis/STA calls & 10 & 25 & 60 \\
Full PnR calls & 3 & 8 & 20 \\
Wall-clock limit & 4 h & 10 h & 24 h \\
\bottomrule
\end{tabular}
\caption{Provisional budget tiers for the paired AXI/DMA tasks.}
\label{tab:budget}
\end{table}

For cumulative cost $c\in[0,B]$, let $r^*(c)$ be the best reward observed by
cost $c$.  We report the normalized area under the progress curve
\begin{equation}
\mathrm{AUC}_B=\frac{1}{B}\int_0^B r^*(c)\,dc,
\end{equation}
alongside final reward, time-to-first-feasible, time-to-threshold, number and
fraction of expensive calls, public--private gap, and run-to-run variance.
A \emph{stagnation episode} is a pre-registered window with no best-so-far
improvement.  Recovery records whether reward subsequently improves and which
layer is edited first.  A cross-stage rollback is an edit to an earlier layer
after later-stage feedback; correctness is evaluated by information gain and
subsequent reward, not by the mere presence of rollback.

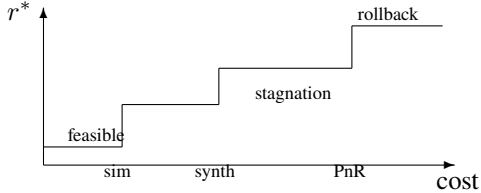
\begin{figure}[t]
\centering
\setlength{\unitlength}{0.8mm}
\begin{picture}(80,34)
  \put(8,5){\vector(1,0){68}}
  \put(8,5){\vector(0,1){26}}
  \put(2,29){$r^*$}
  \put(73,1){cost}
  \put(8,8){\line(1,0){13}}
  \put(21,8){\line(0,1){7}}
  \put(21,15){\line(1,0){16}}
  \put(37,15){\line(0,1){6}}
  \put(37,21){\line(1,0){22}}
  \put(59,21){\line(0,1){7}}
  \put(59,28){\line(1,0){15}}
  \put(18,3){\scriptsize sim}
  \put(33,3){\scriptsize synth}
  \put(56,3){\scriptsize PnR}
  \put(12,9){\scriptsize feasible}
  \put(43,16){\scriptsize stagnation}
  \put(60,29){\scriptsize rollback}
\end{picture}
\caption{A best-so-far trajectory over measured cost. Final reward alone cannot
distinguish early progress, stagnation, and a useful cross-stage rollback after
an expensive PnR call.}
\label{fig:anytime}
\end{figure}

\subsection{Hidden Evaluation and Anti-Gaming}

The final evaluator rebuilds from agent-modified source using pristine scripts.
It varies legal AXI widths, bursts, backpressure, operands, PVT corners, and
floorplan seeds; public and private conditions are disjoint.  RTL and
gate-level simulation prevent simulation--synthesis divergence, LEC catches
logic-changing optimization, an unconstrained-path audit prevents timing
evasion, and DRC/LVS reject physically invalid layouts.  Source manifests lock
testbenches, reference models, tool scripts, and prebuilt binaries outside the
allowed edit surface.  Mutant build failures do not silently reduce a scoring
denominator.

\section{Implementation}

\subsection{Open Toolchain and Harness}

Tasks use Harbor's isolated agent/workspace/verifier layout.  The open toolchain
contains Verilator for simulation, Yosys for synthesis, OpenSTA and OpenROAD
flow scripts for timing and physical implementation, KLayout for layout checks,
and the Sky130HD PDK \cite{ajayi2019openroad}.  The pinned Phase-0 image is
\texttt{openroad/orfs:26Q3-52-gc90beac09} (Yosys 0.64 and KLayout 0.30.7).
Each task ships an agent-visible public loop and a hidden verifier that rebuilds
from a pristine flow and writes structured rewards.  ORFS per-stage JSON is
consumed directly by the trajectory layer.

\subsection{Paired AXI/DMA Vertical Slice}

The shared design, \texttt{firdma}, is a streaming FIR accelerator with an
AXI-Lite control plane, DMA-style read/write engines, configurable bursts and
backpressure, FIFO buffering, interrupts, status and error handling, and an
SRAM macro.  A single specification defines register behavior, FIR arithmetic,
DMA ordering, abort semantics, and performance targets.  The current golden
implementation comprises seven SystemVerilog modules plus a macro wrapper.  It
passes Verilator lint and ten public randomized seeds.  Yosys/slang synthesis of
the initial flop-FIFO version produced 23,789 cells; a 512-entry Sky130 SRAM
macro replacement and physical-flow configuration are present in the current
vertical slice.  At the time of this draft, the macro-based flow has reached
global routing; signoff and task packaging remain pending.

\subsection{Pilot Suite}

Before committing to the vertical slice, we built ten executable tasks spanning
four repositories and seven capability types (Table~\ref{tab:pilot-suite}).
They are not presented as a statistically complete leaderboard.  They validate
the harness, reveal reward-design failure modes, and provide completion--closure
controls that inform the paired protocol.

\begin{table*}[t]
\centering
\small
\begin{tabular}{p{0.21\textwidth}p{0.18\textwidth}p{0.23\textwidth}p{0.28\textwidth}}
\toprule
Task(s) & Capability & Score/oracle & Current evidence and role \\
\midrule
PicoRV32 fmax & RTL timing optimization & hidden RTL+GLS, pristine STA & continuous pilot; target is saturated \\
PicoRV32 coverage & verification closure & Verilator line coverage & iterative pilot; needs mutation/formal upgrade \\
PicoRV32 debug & RTL repair & 6 hidden tests & smoke task; saturated by a frontier agent \\
Vortex SimX DSE & constrained DSE & kernel speedup under area cap & continuous pilot; current cap is saturated \\
Vortex SIMT debug & model debugging & 6 scored kernels & smoke task; several models solve it \\
Vortex dual-model & RTL/reference differential & 3 hidden kernels & validates cross-model oracle; locally solvable \\
Vortex multifault & graded RTL repair & 3 independent faults & one valid Haiku run repairs 2/3 faults \\
AXI DWC repair & protocol/datapath repair & hidden byte scoreboard & matched completion control; saturated \\
AXI DWC verify & mutation-kill closure & 17 hidden faults + golden gate & matched closure control; discriminative \\
Ibex secure abort & security/non-interference & hidden RVFI traces & real upstream defect; agent runs pending \\
\bottomrule
\end{tabular}
\caption{The current pilot suite. ``Saturated'' describes the present instance
and reward cap, not the broader capability.}
\label{tab:pilot-suite}
\end{table*}

\section{Experiments}

\subsection{Experimental Protocol}

Pilot trials use the task-specific wall-clock limits encoded in Harbor and
network-disabled task containers.  We report a score only when the agent starts
and the verifier produces a result.  API, authentication, setup, and network
failures are infrastructure errors.  A timed-out agent is retained when it
leaves a scoreable patch, with the timeout state reported.  Oracle runs validate
score ceilings but are not agent baselines.  The current evidence contains
unequal trial counts and is therefore diagnostic rather than a model ranking.

The central experiment will run the same model/scaffold pairs on Tasks A--C at
all three budgets with three independent trials.  It will report final reward,
AUC, first-feasible time, PnR calls, public--private gap, rollback precision, and
run validity.  This matrix is outside the scope of the present pilot report; a
future benchmark release will freeze model versions, scaffolds, seeds, hardware
allocation, task checksums, and the final signoff oracle before running it.

\subsection{Physical-Flow Feasibility}

We first tested the full ORFS path on a 96-core host.  Table~\ref{tab:orfs}
shows that a 27.6k-instance PicoRV32 design completes in 18.3 minutes and a
492.7k-instance, six-macro design in 22.3 minutes.  Global placement and
detailed routing account for 90\% of the PicoRV32 runtime, supporting separate
budgets for cheap synthesis and expensive PnR.  All three flows report zero
route DRC violations; the PicoRV32 flow closes with $+5.44$ ns WNS at a 12 ns
clock.  These measurements validate the proposed 4/10/24-hour tiers, but do not
yet establish the final \texttt{firdma} difficulty.

\begin{table}[t]
\centering
\small
\begin{tabular}{@{}lrrrl@{}}
\toprule
Design & Instances & Macros & Time & Final status \\
\midrule
GCD & 751 & 0 & 2.9 m & WNS $-1.46$ ns \\
PicoRV32 & 27,590 & 0 & 18.3 m & WNS $+5.44$ ns \\
Chameleon & 492,693 & 6 & 22.3 m & WNS $-0.04$ ns \\
\bottomrule
\end{tabular}
\caption{Phase-0 ORFS runs on Sky130HD. All three generate GDS with zero route
DRC violations.}
\label{tab:orfs}
\end{table}

\subsection{A Completion--Closure Control}

The most informative pilot is a matched pair on the PULP AXI data-width
converter.  In \texttt{axi-dwc-repair}, the agent repairs two byte-lane defects
in a fixed 64-to-128-bit upsizer.  In \texttt{axi-dwc-verify}, the agent may edit
only a Verilator testbench and is scored by the fraction of 17 hidden converter
faults it kills, conditional on no false alarm on golden RTL.  The same three
agent/model configurations were run once on verification and twice on repair.

\begin{table}[t]
\centering
\small
\begin{tabular}{@{}lcc@{}}
\toprule
Agent/model & Repair & Verify \\
\midrule
Codex/GPT-5.6 & 1.000 (2/2) & \best{1.000} (17/17) \\
OpenCode/Big-Pickle & 1.000 (2/2) & 0.0588 (1/17) \\
OpenCode/DeepSeek-V4 & 1.000 (2/2) & 0.000 (golden alarm) \\
\bottomrule
\end{tabular}
\caption{Matched AXI completion and verification-closure tasks. Parentheses are
successful repair trials or killed hidden faults.}
\label{tab:axi}
\end{table}

All agents solve the localized repair, so additional seeds would not make that
instance a meaningful long-horizon test.  The open-ended verification task
separates them sharply: Codex kills all faults in 10.6 minutes of agent
execution, Big-Pickle kills one in 28.7 minutes, and DeepSeek false-alarms on
golden behavior after 26.7 minutes.  This controlled contrast supports the
benchmark's completion--closure premise, although single verification trials
do not support general model-ranking claims.

\subsection{What the Remaining Pilots Reveal}

The fmax task shows why raw metrics must accompany normalized reward: Codex
reaches 181.7 MHz against a 120 MHz full-score target, while the oracle reaches
99.0 MHz and Big-Pickle 110.1 MHz.  The reward cap therefore discards meaningful
variation.  Likewise, Codex reaches 88.7\% line coverage against an 85\% cap;
line coverage alone does not establish checker strength.  Vortex SIMT and
dual-model debugging are solved by multiple agents, whereas the three-fault
Vortex task yields a graded 2/3 on its valid Haiku run.  These cases motivate
four requirements adopted by \closer{}: hidden condition changes rather than
only new seeds, graded independent objectives, target calibration above current
frontier performance, and run-status provenance.

\subsection{Scope of the Current Evaluation}

The current evidence establishes the executable harness, physical-flow
feasibility, and the completion--closure contrast; it does not yet rank models
on the full stage-paired protocol.  That evaluation requires paired A$+$B
versus C runs at equal aggregate budget (RQ1), recovery confusion matrices over
RTL/SDC/floorplan/flow layers (RQ2), and low/medium/high progress curves with
bootstrap confidence intervals (RQ3).  We defer these analyses until the final
signoff oracle and repeated-trial matrix are frozen.

\section{Discussion and Limitations}

\paragraph{What \closer{} measures.}
The intended object is not chip-design knowledge in isolation.  It is a policy
for allocating expensive experiments under delayed, cross-layer feedback.  A
model may know how to fix RTL and tune OpenROAD separately yet fail end-to-end
because it overuses PnR, trusts a misleading proxy, or never revisits an RTL
decision.  Stage pairing exposes that gap without claiming to be the first
RTL-to-GDS or specification-to-GDS system.

\paragraph{Current limitations.}
The stage-paired vertical slice currently contains one primary design, one PDK,
and an incomplete final signoff package.  The pilot suite is heterogeneous and
has unequal model coverage; many tasks have only one valid trial and several
are saturated.  Manually injected faults do not substitute for real historical
bug clusters.  Open-source Sky130 flows improve reproducibility but do not
represent commercial signoff.  Token counts are not directly comparable across
agent scaffolds, and wall-clock measurements depend on host contention.

A complete stage-paired model comparison additionally requires the macro-based
\texttt{firdma} flow to close through DRC/LVS and multi-corner STA, Tasks A--C
and the trajectory schema to be frozen, and every model/task/budget cell to
contain repeated valid trials.  The corresponding artifact should include task
checksums, containers, raw trajectories, and an explicit invalid-run manifest.

\section{Conclusion}

\closer{} reframes hardware-agent evaluation around a controlled question:
given the same design objective, does an agent benefit from explicit stage
boundaries, and can it spend limited EDA calls and recover across those
boundaries when they are removed?  Its paired \specToRTL{}, \rtlToGDS{}, and
\specToGDS{} tasks share intent and hidden conditions, while its trajectory
protocol makes final quality, anytime progress, cost, and rollback observable.
The completed pilot suite and AXI completion--closure contrast show why final
pass/fail and runtime alone are insufficient.  The remaining central experiment
will determine whether the proposed protocol yields stable, model-discriminative
evidence at full physical-design depth.

\bibliography{references}

@misc{liu2023verilogeval,
  author = {Mingjie Liu and Nathaniel Pinckney and Brucek Khailany and Haoxing Ren},
  title = {{VerilogEval}: Evaluating Large Language Models for Verilog Code Generation},
  year = {2023},
  eprint = {2309.07544},
  archivePrefix = {arXiv},
  url = {https://arxiv.org/abs/2309.07544}
}

@inproceedings{lu2024rtllm,
  author = {Yao Lu and Shang Liu and Qijun Zhang and Zhiyao Xie},
  title = {{RTLLM}: An Open-Source Benchmark for Design {RTL} Generation with Large Language Models},
  booktitle = {Proceedings of the Asia and South Pacific Design Automation Conference},
  year = {2024},
  doi = {10.1109/ASP-DAC58780.2024.10473904}
}

@misc{liu2025openllmrtl,
  author = {Shang Liu and Yao Lu and Wenji Fang and Mengming Li and Zhiyao Xie},
  title = {{OpenLLM-RTL}: Open Dataset and Benchmark for {LLM}-Aided Design {RTL} Generation},
  year = {2025},
  eprint = {2503.15112},
  archivePrefix = {arXiv},
  url = {https://arxiv.org/abs/2503.15112}
}

@misc{pinckney2025cvdp,
  author = {Nathaniel Pinckney and Chenhui Deng and Chia-Tung Ho and Yun-Da Tsai and Mingjie Liu and Wenfei Zhou and Brucek Khailany and Haoxing Ren},
  title = {Comprehensive Verilog Design Problems: A Next-Generation Benchmark Dataset for Evaluating Large Language Models and Agents on {RTL} Design and Verification},
  year = {2025},
  eprint = {2506.14074},
  archivePrefix = {arXiv},
  url = {https://arxiv.org/abs/2506.14074}
}

@misc{yu2026chipbench,
  author = {Zhongkai Yu and Chenyang Zhou and Yichen Lin and Hejia Zhang and Haotian Ye and Junxia Cui and Zaifeng Pan and Jishen Zhao and Yufei Ding},
  title = {{ChipBench}: A Next-Step Benchmark for Evaluating {LLM} Performance in {AI}-Aided Chip Design},
  year = {2026},
  eprint = {2601.21448},
  archivePrefix = {arXiv},
  url = {https://arxiv.org/abs/2601.21448}
}

@misc{cui2026hwe,
  author = {Fan Cui and Hongyuan Hou and Zizhang Luo and Chenyun Yin and Yun Liang},
  title = {{HWE-Bench}: Benchmarking {LLM} Agents on Real-World Hardware Bug Repair Tasks},
  year = {2026},
  eprint = {2604.14709},
  archivePrefix = {arXiv},
  url = {https://arxiv.org/abs/2604.14709}
}

@misc{zou2026phoenix,
  author = {Qingyun Zou and Feng Yu and Hongshi Tan and Bingsheng He and WengFai Wong},
  title = {Is Agentic {AI} Ready for Real-World Hardware Engineering? A Deep Dive with {Phoenix-bench}},
  year = {2026},
  eprint = {2605.15226},
  archivePrefix = {arXiv},
  url = {https://arxiv.org/abs/2605.15226}
}

@misc{yu2026horizon,
  author = {Cunxi Yu and Chenhui Deng and Nathaniel Pinckney and Brucek Khailany},
  title = {Agentic Hardware Design as Repository-Level Code Evolution},
  year = {2026},
  eprint = {2606.28279},
  archivePrefix = {arXiv},
  url = {https://arxiv.org/abs/2606.28279}
}

@article{he2024chateda,
  author = {Zhuolun He and Haoyuan Wu and Xinyun Zhang and Xufeng Yao and Su Zheng and Haisheng Zheng and Bei Yu},
  title = {{ChatEDA}: A Large Language Model Powered Autonomous Agent for {EDA}},
  journal = {IEEE Transactions on Computer-Aided Design of Integrated Circuits and Systems},
  year = {2024},
  doi = {10.1109/TCAD.2024.3383347}
}

@misc{ghose2025orfs,
  author = {Amur Ghose and Andrew B. Kahng and Sayak Kundu and Zhiang Wang},
  title = {{ORFS-agent}: Tool-Using Agents for Chip Design Optimization},
  year = {2025},
  eprint = {2506.08332},
  archivePrefix = {arXiv},
  url = {https://arxiv.org/abs/2506.08332}
}

@inproceedings{allam2025asic,
  author = {Ahmed Allam and Youssef Mansour and Mohamed Shalan},
  title = {{ASIC-Agent}: An Autonomous Multi-Agent System for {ASIC} Design with Benchmark Evaluation},
  booktitle = {IEEE International Conference on LLM-Aided Design},
  year = {2025},
  doi = {10.1109/ICLAD65226.2025.00033}
}

@misc{shi2026cktevo,
  author = {Zhengyuan Shi and Jingxin Wang and Tairan Cheng and Changran Xu and Weikang Qian and Qiang Xu},
  title = {{CktEvo}: Repository-Level {RTL} Code Benchmark for Design Evolution},
  year = {2026},
  eprint = {2603.08718},
  archivePrefix = {arXiv},
  url = {https://arxiv.org/abs/2603.08718}
}

@misc{li2026pdagent,
  author = {Qiufeng Li and Rongqian Chen and Quan Cheng and Chengxuan Wang and Sizhe Tang and Wuxi Li and Duo Ding and Chia-Tung Ho and Haoxing Ren and David Z. Pan and Tian Lan and Weidong Cao},
  title = {{PDAGENT-BENCH}: Characterizing, Grounding, and Architecting {LLM} Agents for {VLSI} Physical Design},
  year = {2026},
  eprint = {2606.17253},
  archivePrefix = {arXiv},
  url = {https://arxiv.org/abs/2606.17253}
}

@misc{verkor2026conductor,
  author = {{The Verkor Team} and Ravi Krishna and Suresh Krishna and David Chin},
  title = {Design Conductor: An Agent Autonomously Builds a 1.5 {GHz} Linux-Capable {RISC-V} {CPU}},
  year = {2026},
  eprint = {2603.08716},
  archivePrefix = {arXiv},
  url = {https://arxiv.org/abs/2603.08716}
}

@misc{goldblatt2026hightide,
  author = {Benjamin Goldblatt and Paolo Pedroso and Farhad Modaresi and Ethan Sifferman and Matthew R. Guthaus},
  title = {{HighTide}: An Agent-Curated Open-Source {VLSI} Benchmark Suite},
  year = {2026},
  eprint = {2606.04126},
  archivePrefix = {arXiv},
  url = {https://arxiv.org/abs/2606.04126}
}

@misc{tsai2026hsco,
  author = {Pei-Huan Tsai and Kuan-Lin Chiu and William Baisi and Pin-Yu Chen and Luca P. Carloni},
  title = {{HSCO-Bench}: An Agent-Driven End-to-End Hardware-Software Co-design Benchmark for Systems-on-Chip},
  year = {2026},
  eprint = {2605.19399},
  archivePrefix = {arXiv},
  url = {https://arxiv.org/abs/2605.19399}
}

@inproceedings{ajayi2019openroad,
  author = {Tutu Ajayi and Vidya A. Chhabria and Mateus Foga{\c{c}}a and Soheil Hashemi and Abdelrahman Hosny and Andrew B. Kahng and Minsoo Kim and Jeongsup Lee and Uday Mallappa and Marina Neseem and Geraldo Pradipta and Sherief Reda and Mehdi Saligane and Sachin S. Sapatnekar and Carl Sechen and Mohamed Shalan and William Swartz and Lutong Wang and Mingyu Woo and Bang Xu},
  title = {Toward an Open-Source Digital Flow: First Learnings from the {OpenROAD} Project},
  booktitle = {Proceedings of the 56th Annual Design Automation Conference},
  year = {2019},
  doi = {10.1145/3316781.3326334}
}

\end{document}